\documentclass[aps,prb,twocolumn,groupedaddress]{revtex4}
\usepackage{ae,aecompl}
\usepackage{helvet}
\usepackage{graphicx}
\usepackage{dcolumn}
\usepackage{color}
\usepackage[latin9]{inputenc}
\usepackage{amssymb}
\usepackage{amsmath}

\def\be{\begin{equation}}
\def\ee{\end{equation}}
\def\br{\mathbf{r}}
\newcommand{\eps}{\varepsilon}
\newcommand{\corr}[1]{\langle #1\rangle}

\begin{document}
\bibliographystyle{apsrev}


\title{Spectroscopic evidence for strong correlations between local superconducting gap and local Altshuler-Aronov density-of-states suppression in ultrathin NbN films}

\author{C. Carbillet$^{1}$}
\author{V. Cherkez$^{1}$}
\author{M.A. Skvortsov$^{2,3}$}
\email{skvor@itp.ac.ru}
\author{M.V. Feigel'man$^{3,2}$}
\author{F. Debontridder$^{1}$}
\author{L.B. Ioffe$^{4,3}$}
\author{V.S. Stolyarov$^{1,5,6}$}
\author{K. Ilin$^{7}$}
\author{M. Siegel$^{7}$}
\author{C. No\^us$^{8}$}
\author{D. Roditchev$^{1,9}$}
\author{T. Cren$^{1}$}
\author{C. Brun$^{1}$}
\email{christophe.brun@sorbonne-universite.fr}

\affiliation{$^1$Sorbonne Universit\'{e}, CNRS, Institut des Nanosciences de Paris, UMR7588, F-75252, Paris, France}
\affiliation{$^2$Skolkovo Institute of Science and Technology, Moscow 121205, Russia}
\affiliation{$^3$L. D. Landau Institute for Theoretical Physics, Chernogolovka 142432, Russia}
\affiliation{$^4$Sorbonne Universit\'{e}, CNRS, LPTHE, UMR 7589, F-75252, Paris, France}
\affiliation{$^5$Moscow Institute of Physics and Technology, 141700 Dolgoprudny, Russia}
\affiliation{$^6$Dukhov Research Institute of Automatics (VNIIA), 127055 Moscow, Russia}
\affiliation{$^7$Institute of Micro- und Nano-electronic Systems, Karlsruhe Institute of Technology, Hertzstrasse 16, D-76187 Karlsruhe, Germany}
\affiliation{$^8$Sorbonne Universit\'{e}, CNRS, Laboratoire Cogitamus, UMR7588, F-75252, Paris, France}
\affiliation{$^9$LPEM, ESPCI Paris-PSL Research University-Sorbonne Universit\'{e}, F-75005 Paris, France}

\date{\today}

\begin{abstract}
Disorder has different profound effects on superconducting thin films. For a large variety of materials, increasing disorder reduces electronic screening which enhances electron-electron repulsion. These fermionic effects lead to a mechanism described by Finkelstein: when disorder combined to electron-electron interactions increases, there is a global decrease of the superconducting energy gap $\Delta$ and of the critical temperature $T_c$, the ratio $\Delta$/$k_BT_c$ remaining roughly constant. In addition, in most films an emergent granularity develops with increasing disorder and results in the formation of inhomogeneous superconducting puddles. These gap inhomogeneities are usually accompanied by the development of bosonic features: a pseudogap develops above the critical temperature $T_c$ and the energy gap $\Delta$ starts decoupling from $T_c$.
Thus the mechanism(s) driving the appearance of these gap inhomogeneities could result from a complicated interplay between fermionic and bosonic effects. By studying the local electronic properties of a NbN film with scanning tunneling spectroscopy (STS) we show that the inhomogeneous spatial distribution of $\Delta$ is locally strongly correlated to a large depletion in the local density of states (LDOS) around the Fermi level, associated to the Altshuler-Aronov effect induced by strong electronic interactions. By modelling quantitatively the measured LDOS suppression, we show that the latter can be interpreted as local variations of the film resistivity. This local change in resistivity leads to a local variation of $\Delta$ through a local Finkelstein mechanism. Our analysis furnishes a purely fermionic scenario explaining quantitatively the emergent superconducting inhomogeneities, while the precise origin of the latter remained unclear up to now.
\end{abstract}

\maketitle

\section{Introduction}

Non-magnetic disorder was initially thought to have little effect on the superconducting properties of conventional thin films, as far as time-reversal symetry is preserved \cite{Anderson,Abrikosov}. Decades of intense experimental and theoretical works have ended to the opposite conclusion showing that beyond a critical disorder all systems transit either to a metallic or to an insulating state \cite{Gantmakher,Feigelman}. For thin films that are not single-crystals \cite{Brun2017} but consist of coupled nanocrystals two different classes of systems were historically considered. A qualitatively different superconducting behavior was observed between so-called \emph{granular} and \emph{homogeneous} disordered thin films. In \emph{granular} ones a poor electrical coupling between the nanocrystals makes the film a disordered array of Josephson junctions \cite{Efetov,Orr,White,Jaeger,Gantmakher}. Their superconducting properties are controlled by
weak Josephson couplings between phases of local superconducting order parameters attributed to individual nanocrystals.
In \emph{homogeneous} ones the nanocrystals are much better electrically coupled to each other. Their superconducting properties are controlled by the subtle interplay between the non-magnetic disorder distribution and electron-electron interactions \cite{Graybeal,Dynes,Haviland,Feigelman}.

Nevertheless experimental and theoretical studies performed in the last ten years enabled to show that this historical categorization between \emph{granular} and \emph{homogeneous} thin films is lacking of an important link between them. At low disorder in \emph{homogeneously disordered} films macroscopic measurements show that $T_c$ and $\Delta$ monotonously decrease with rising disorder corresponding to $k_F  l$ decreasing from a value much larger than one toward unity \cite{Graybeal,Dynes,Haviland,Feigelman} ($k_F$ being the Fermi wavevector and $l$ the elastic electron mean-free path). This effect has been satisfactorily explained by a reduction of the electronic screening upon increasing disorder which reduces the diffusion constant $D$ and is called the ``fermionic'' mechanism \cite{Finkelstein,Finkelstein1994}. Many films like, for example, Bi, Pb, MoGe, NbSi behave this way and follow the so-called Finkelstein scenario down to the almost complete destruction of superconductivity~\cite{Dynes,Valles,Graybeal,Crauste1,Crauste2}.

However a different behavior can also take place for sufficiently large disorder in nominally \emph{homogeneous} films, i.e.\ typically for $1 \le k_F l\le 3$: in amorphous InO$_x$, TiN or NbN the single-particle gap probed by tunneling remains finite and decouples from the decreasing $T_c$, while a pseudogap emerges in a significant temperature range above $T_c$ \cite{Sacepe2008,Sacepe2010,Sacepe2011,Chand2012}. These latter features are believed to correspond to the so-called ``bosonic'' scenario, where disorder localizes single-electron wavefunctions and provoke an ``emergent granularity" in the superconductor \cite{Ghosal,Feigelman,Bouadim}. These emergent superconducting puddles were revealed by scanning tunneling microscopy/spectroscopy (STM/STS) in TiN, InO$_x$ and NbN \cite{Sacepe2008,Sacepe2011,Chand2012,Noat,Kamlapure}. From moderate to strong disorder, the long-range phase coherence between the emergent neighboring superconducting islands gradually weakens and eventually destroys, leaving the material with residual incoherent pairing properties \cite{Feigel2007,Feigel2010}.

\begin{figure}
\includegraphics[width=0.8\columnwidth]{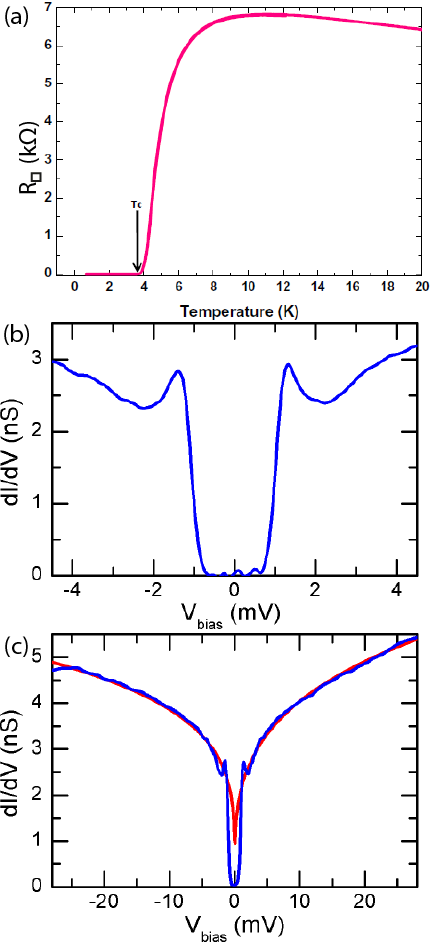}
\caption{\small (Color online) (a) Square resistance of a 2.1~nm thick NbN film, measured in the same stage where STM is performed. $T_c$ is defined when $R_\Box$ reaches zero, leading to $T_c=3.8$~K. (b) Typical local $dI/dV(V)$ spectrum measured at $T=300$~mK.
Set-point for spectroscopy $V=-10$~mV, $I=150$~pA. (c) Local $dI/dV(V)$ spectrum measured at the same location and temperature as in (b) but on a larger voltage scale.
The large depletion of the LDOS seen around $E_F$ has a characteristic V-shape due to electron-electron interactions enhanced by disorder. The red (lighter) curve shows a {power-law} fit of this V-shape dependence, according to the relation $dI/dV(V)=b \times V^{\alpha}$ where $b$ is a constant.} \label{Fig1}
\end{figure}

From the above considerations one sees that in fact most films are expected to present an interplay between ``fermionic'' and ``bosonic'' effects for strong enough disorder before superconductivity disappears. Nevertheless there is a lack of experimental data combining macroscopic and local measurements on the same system to carefully explore this interplay. Such a detailed study was conducted in NbN~\cite{Chand2012} and enabled to draw a phase diagram, where at low disorder the fermionic mechanism dominates, while bosonic effects develop at stronger disorder, as exposed above.

In this work we show that, in contrast to the common belief, the emergent gap inhomogeneities can be solely explained by the Finkelstein mechanism, where electron interactions play a crucial role. Our claim is supported by the fact that using scanning tunneling microscopy/spectroscopy in NbN thin films, we found a direct spatial cross-correlation between local superconducting gap maps and the mapping of the Altshuler-Aronov suppression of the local density-of-states (LDOS). Through appropriate modeling we demonstrate that both effects can be linked to local variations of the film's resistance. Due to an interplay between disorder and electrons interaction, these resistivity variations drive corresponding spatial variations of the normal-state STS spectra and the local energy gap values.



\color{black}

\begin{figure*}
\includegraphics[width=0.9\textwidth]{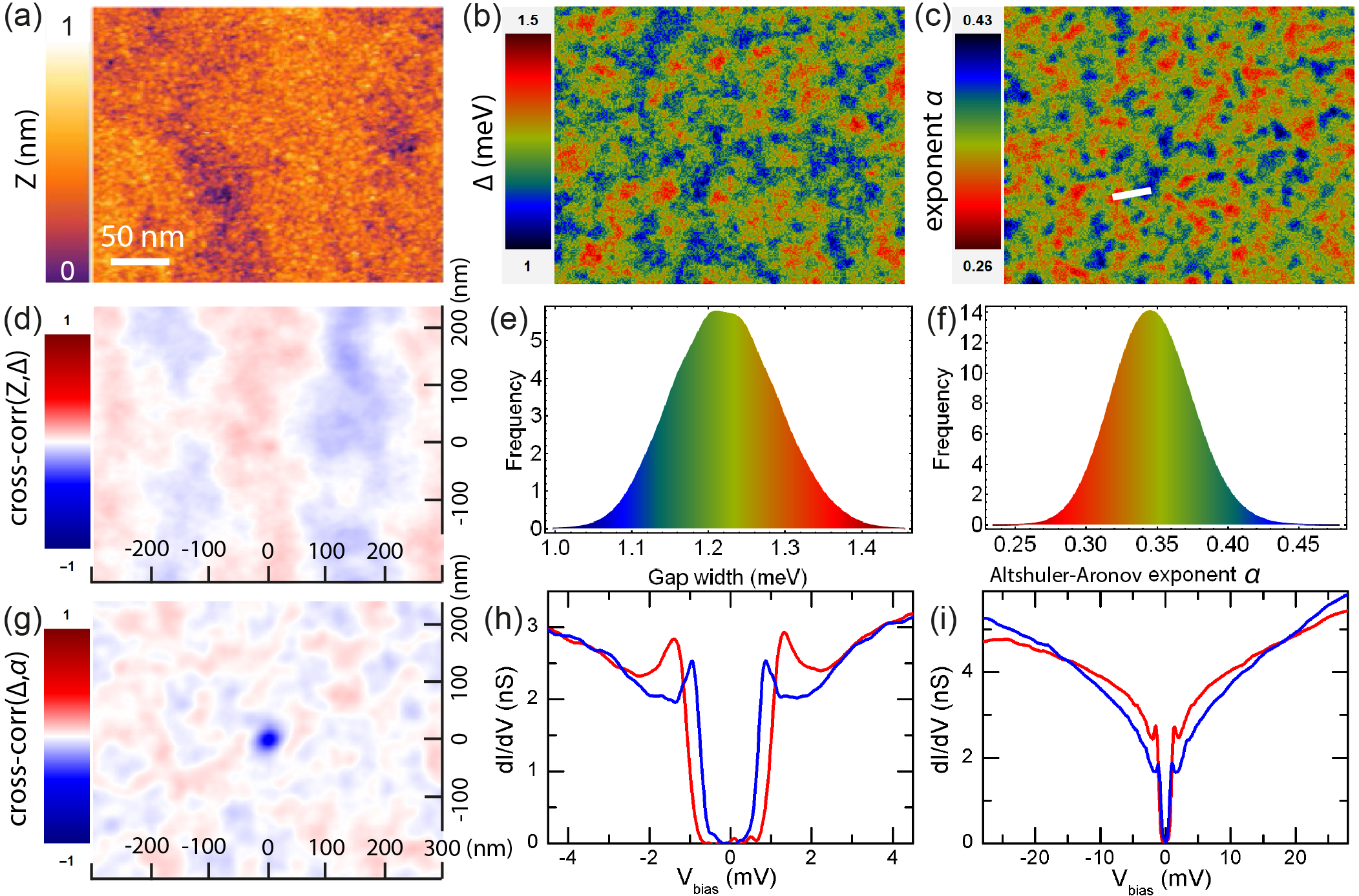}
\caption{(Color online) Spatial cross-correlations between variations of local the superconducting gap $\Delta(\mathbf{r})$ and the Altshuler-Aronov exponent $\alpha(\mathbf{r})$. Measurements done at 300~mK in zero magnetic field. (a) STM topography of a $300\times250$~nm$^2$ area of the NbN sample (height variation $Z$ in nm). Scanning parameters: $V=-50$~mV and $I=50$~pA. (b) Color-coded map showing the local energy gap $\Delta(\mathbf{r})$ (in meV) measured in area (a) by STS. (c) Color-coded map showing the local exponent $\alpha(\mathbf{r})$ in area (a). (d) Color-coded map showing the cross-correlation between the topography $Z(x,y)$ shown in (a) and the gap map $\Delta(x,y)$ shown in (b). No cross-correlations are found. (e) Histogram of $\Delta(\mathbf{r})$ values occuring in (b). (f) Histogram of $\alpha(\mathbf{r})$ values occuring in (c). (g) Color-coded map showing the cross-correlation between the gap map $\Delta(x,y)$ shown in (b) and the exponent map $\alpha(x,y)$ shown in (c). A strong spatial anti-correlation is found: where local electron-electron interactions are stronger (larger $\alpha(x,y)$) $\Delta(x,y)$ is smaller and vice-versa. (h): Representative $dI/dV$ spectra of the regions seen in (b). Brighter (respectively darker) spectra are measured in brighter (darker) regions of panel (b). (i) Same spectra as in (h) but over an energy scale 20 to 30 times larger than $\Delta$. Brighter (respectively darker) spectra are measured in brighter (darker) regions of panel (c).} \label{Fig2}
\end{figure*}

\section{Experimental results: local STS measurements of the gap and V-shape LDOS}

Our samples consist of ultrathin NbN films grown ex-situ on sapphire, structured in nano-crystals of lateral size 2--5~nm \cite{Semenov}. Special care was taken in order to minimize the thickness of the surface oxyde layer: the freshly grown samples were introduced into our STM set-up in less than 24 hours after their growth. We selected films having a $T_c \approx 0.25 \, T_c^\text{bulk} = 3.8$~K, so that seeming ``bosonic'' properties like gap inhomogeneities and pseudogap features have already developed. The nominal film thickness is about 2.1~nm. We studied their local superconducting properties by scanning tunneling microscopy/spectroscopy (STM/STS) at 300~mK, using PtIr tips, in an ultrahigh vacuum homemade setup. The presented $dI/dV$ measurements were obtained by numerical derivation of single $I(V)$ curves. $T_c$ was extracted from \emph{in situ} 4-points electrical resistivity mesurements performed in the STM stage during the same run as the STM/STS measurements. The temperature dependence of the resitivity before the superconducting transition is typical of 2D disordered superconducting films and is shown in Fig.\ \ref{Fig1}a.

STM topography measurements show that the film surface is very flat (see Fig.\ \ref{Fig2}a). The size of the observed nanoscale structures correlates well with the size of the nanocrystals characterized independently by transmission electron microscopy (TEM) experiments. A characteristic $dI/dV$ spectrum measured locally is shown in Fig.\ \ref{Fig1}b. It presents a fully gapped LDOS with well-defined superconducting coherence peaks. We also note that in-gap states are also present, most probably induced by paramagnetic defects at the NbN/sapphire interface. For energies larger than the ones of the coherence peaks, a strong background is seen instead of recovering a standard normal DOS, as in the Bardeen-Cooper-Schrieffer (BCS) case. Fig.\ \ref{Fig1}c enables characterizing this background: it presents a $dI/dV$ spectrum measured on the same location but over a wider energy range about 20 times larger than $\Delta$. We see that the tunneling LDOS is strongly reduced over an energy range much larger than $\Delta$. This
LDOS suppression is typical of systems where electronic correlations combined to disorder reduce the tunneling DOS at the Fermi level ($E_F$) \cite{Altshuler,Altshuler1985}, an effect first theoretically explained by Altshuler and Aronov. In the following we refer to this effect as AA.

Fig.\ \ref{Fig1}c shows that for $e|V|>\Delta$ the measured LDOS follows a power-law as a function of the applied bias voltage, $dI/dV\propto V^{\alpha}$. Please note that here the fit is truly performed using the functional dependence $V^{\alpha}$ without any offset: this curve goes to zero at zero voltage. Nevertheless since the fit is performed in the energy region outside the gap, data points exactly around zero voltages are not plotted in this figure. Such a behavior has also being evidenced for instance in highly disordered metallic thin films close to the insulating transition \cite{Richardella}.

\begin{figure*}
\includegraphics[width=\textwidth]{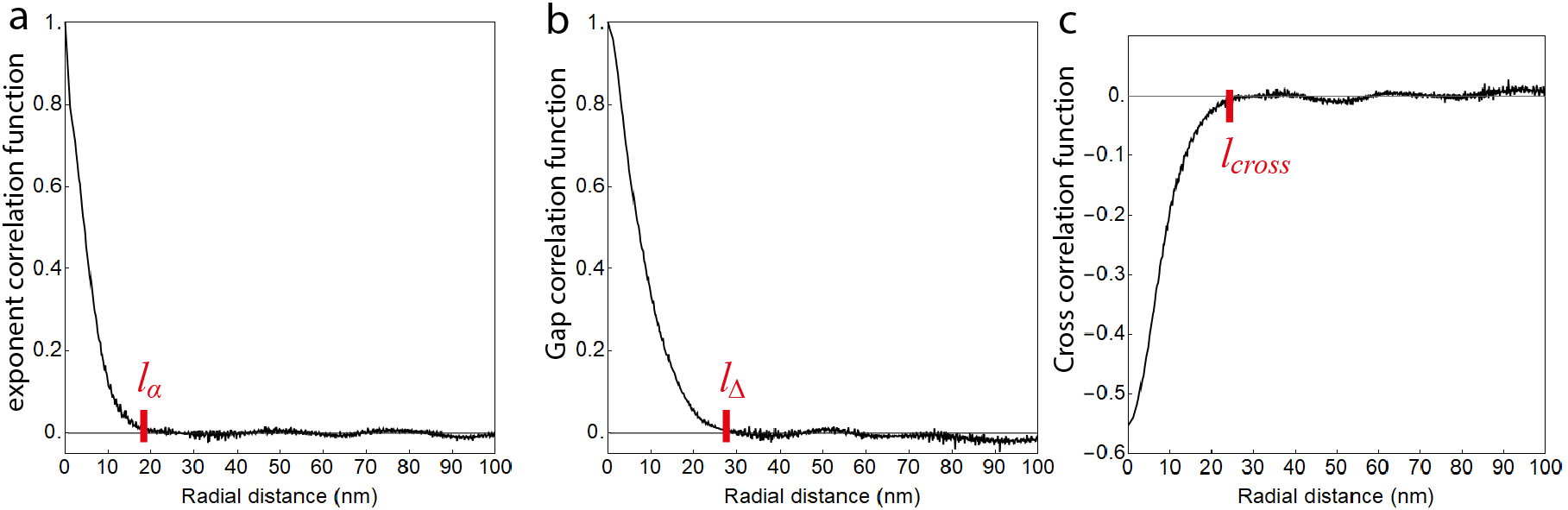}
\caption{(Color online) Spatial decay of the correlation functions calculated from the radial averaging of the 2D auto-correlation or cross-correlation functions from the data shown in Fig.~\ref{Fig2}. Each profile starts at the center of each map (corresponding to the origin 0) and extends up to 100~nm. (a) Spatial dependence of $\rho_{\alpha}(r)$ defined as the correlation function of the Altshuler-Aronov exponent $\alpha$(r). (b) Spatial dependence of $\rho_{\Delta}(r)$ defined as the correlation function of the superconducting energy gap $\Delta$(r). (c) Spatial dependence of $\rho_\text{cross}(r)$ defined as the correlation function of the 2D cross-correlation between the local Altshuler-aronov exponent $\alpha(\mathbf{r})$ and superconducting energy gap $\Delta(\mathbf{r})$.} \label{Fig3}
\end{figure*}

Theoretically such a dependence is expected in the framework of nonperturbative extension~\cite{Ingold,LevitovShytov,Rollbuhler} of the Altshuler-Aronov theory~\cite{Altshuler,Altshuler1985}. Let us note that this kind of effect is also referred to as dynamical Coulomb blockade effect in mesoscopic physics~\cite{Ingold,LevitovShytov,Rollbuhler}. Interestingly, it was shown that the effect of probing the tunneling DOS of an ultrasmall tunnel junction in a 2D system with electron-electron interaction is equivalent to dynamical Coulomb blockade formalism with an ohmic environment\cite{Rollbuhler}. We have already used such an equivalence when modeling the superconducting proximity effect between a Pb single nano-crystal and a 2D metallic homogeneously disordered Pb film~\cite{Brun2012,Serrier,Roditchev}.

The main finding of our work is a systematic spatial correlation between the local energy gap value $\Delta(\mathbf{r})$ and the strength of the AA LDOS suppression, quantified by the exponent $\alpha(\mathbf{r})$ in the power law $dI/dV(\mathbf{r})\propto V^{\alpha(\mathbf{r})}$.
At each square nanometer of the $300\times250$~nm$^2$ area shown in Fig.~\ref{Fig2}a, the local gap value was extracted from the peak-to-peak distance in each $dI/dV(\mathbf{r})$ curve. The extracted map of the gap values is shown in Fig.~\ref{Fig2}b. The energy gap ranges between 1 and about 1.5~meV. The histogram of the gap values is shown in panel~\ref{Fig2}e and is close to a Gaussian distribution. The spatial distribution of $\Delta(\mathbf{r})$ shows emerging local inhomogeneities seen as lighter and darker puddles. Using the scale bar shown in the topography Fig.~\ref{Fig2}a, one sees that local puddles having a constant gap identified by a certain local color are smaller than 50~nm. Such a behavior was reported previously in similar systems for comparable disorder strength, but with comparatively less clean and less dense STS spectra\cite{Sacepe2008,Noat,Kamlapure,Carbillet}. $\Delta = 1.2$ meV is used below as the average gap value of our film.

We call \emph{supergrain} a local puddle with a nearly constant gap. We define $l_{\Delta}$ as the characteristic size of such supergrain. We found $l_{\Delta} \approx 27$ nm which is much larger than the low-$T$ coherence length $\xi(0)\approx 5$ nm. $l_{\Delta}$ corresponds to a correlation length characterizing the spatial extent over which local gap values are correlated. A precise way to extract it is to use the spatial dependence of the radial averaging $\rho_{\Delta}(r)$ of the 2D autocorrelation function of $\Delta(\mathbf{r})$ (details about the numerical computations of the correlation and cross-correlation functions can be found in the Appendices section). The spatial decay of $\rho_{\Delta}(r)$ and $l_{\Delta}$ are shown in Fig.~\ref{Fig3}b.

We provide now for the first time a local mapping of the disorder distribution combined to electron interactions which together are responsible for the spatial distribution of the gap inhomogeneities.
At each location the gap value $\Delta(\mathbf{r})$ was measured, the local AA exponent $\alpha(\mathbf{r})$ was also extracted in the energy range [5;30]~mV, separately for positive and negative energies. We then took for $\alpha(\mathbf{r})$ the average between the values fitted for positive and negative energies. The slight asymmetry in our spectra between positive and negative energies probably comes from a non-constant tip DOS.

\begin{figure}
\includegraphics[width=\columnwidth]{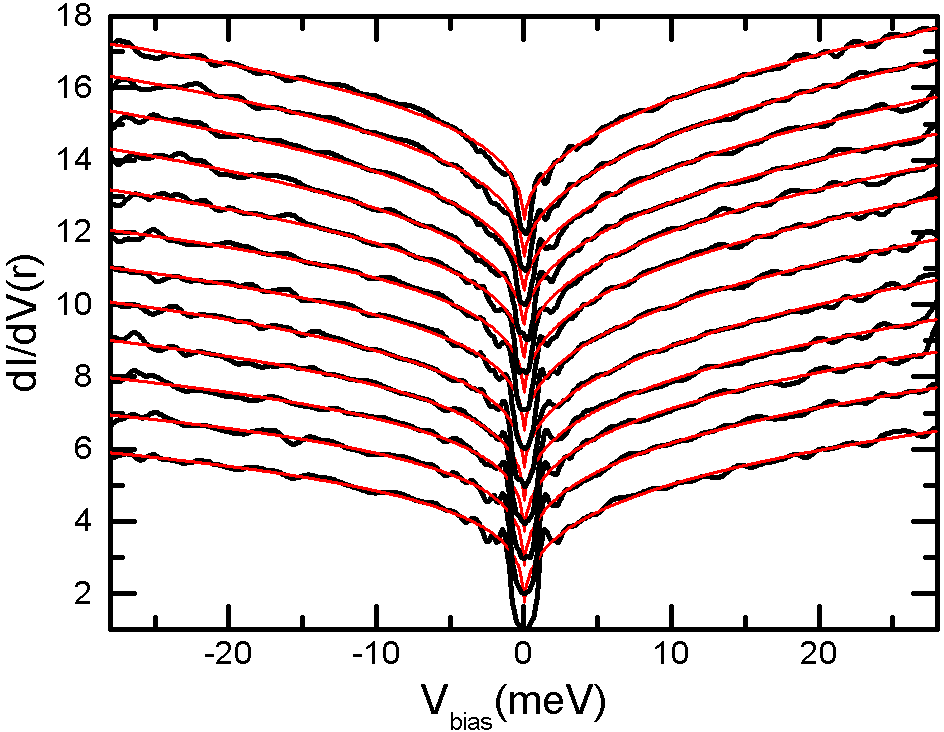}
\caption{\small (Color online) Linecut of twelve $dI/dV(\mathbf{r})$ spectra (shown by black curves) measured along the line indicated by a white rectangle in Fig.~\ref{Fig2}c. This linecut starts at a bright puddle with larger local gap and smaller exponent. 33~nm away it ends at a dark puddle with smaller gap and larger exponent. Each spectrum is spaced by 3~nm from the previous one.  Vertically each spectrum is offset by one for clarity including the first spectrum which is the bottom one. Each $dI/dV(\mathbf{r})$ curve is superposed with its fitted curve (shown by red lines) using a power law $b \times V^{\alpha(\mathbf{r})}$. Set-point for spectroscopy $V=-10$~mV, $I=150$~pA.} \label{Fig4}
\end{figure}

The 2D map of the exponent $\alpha(\mathbf{r})$ is presented in panel \ref{Fig2}c and its histogram in panel~\ref{Fig2}f. Most of the values are situated between 0.26 and 0.43. It is seen that the exponent map also presents emerging granularity, with some resemblance with the gap map. Nevertheless the grainy structure of the exponent map looks sharper than the one of the gap map. This suggests that the correlation length $l_{\alpha}$ of $\alpha(\mathbf{r})$ should be smaller than $l_{\Delta}$. The correlation length $l_{\alpha}$ is extracted following a similar procedure as for $\Delta(r)$ above. The spatial dependence of the radial averaging $\rho_{\alpha}(r)$ of the 2D autocorrelation function of $\alpha(\mathbf{r})$ is shown in Fig.~\ref{Fig3}a. $l_{\alpha}$ also is indicated and yields $l_{\alpha}\approx18$ nm. This value is indeed about two thirds of $l_{\Delta}$. In order to simplify the visual comparison between the gap map and the exponent map, we used the same color scale for both maps but with a reversed order, which makes it easier to see anti-correlations. This means that if statistically anti-correlations dominate between $\Delta(\mathbf{r})$ and $\alpha(\mathbf{r})$ one expects to see at the same place $\mathbf{r}$ bright puddles in both maps: where a large gap exists (bright puddle) a low $\alpha$ value should also be found (bright puddle). The situation is similar for darker puddles. Looking precisely at both maps Fig.~\ref{Fig2}b and Fig.~\ref{Fig2}c and trying to compare the shape and color of local puddles, one sees that anti-correlations are indeed present.

In order to mathematically quantify the cross-correlations existing between $\Delta(\mathbf{r})$ and $\alpha(\mathbf{r})$ we computed their 2D normalized cross-correlation function, which is shown in panel~\ref{Fig2}g. A very large anti-correlation of about $-0.55$ at coincident points is found.
This confirms that the larger is the local energy gap value $\Delta(\mathbf{r})$, the smaller is the local AA exponent $\alpha(\mathbf{r})$ and vice-versa, as illustrated on panels~\ref{Fig2}h and i. The $dI/dV(\mathbf{r})$ spectrum with a large $\Delta(\mathbf{r})$ has a flatter AA background (red curve), and vice versa (blue curve). To extract the spatial extent over which local cross-correlations exist, we calculated the radial averaging $\rho_\text{cross}(r)$ of the 2D cross-correlation function presented in Fig.~\ref{Fig2}g. Its spatial dependence is shown on panel Fig.~\ref{Fig3}c. One sees that the typical correlation length of the cross-correlations is $l_\text{cross}\approx25$~nm. This is larger than $l_{\alpha}$ and close to $l_{\Delta}$.

To show the robustness of our results we present in Fig.~\ref{Fig4} a linecut of twelve $dI/dV(\mathbf{r})$ spectra measured along the line indicated by a white rectangle in Fig.~\ref{Fig2}c. This linecut starts at a bright puddle with larger local gap and smaller exponent. It corresponds to the bottom $dI/dV$ spectrum. 33~nm away this linecut ends at a dark puddle with smaller gap and larger exponent. The end corresponds to the top $dI/dV$ spectrum. Each spectrum is spaced by 3~nm from the previous one. It is seen that each individual $dI/dV(\mathbf{r})$ curve is well-fitted by a power law $dI/dV(\mathbf{r})=b \times V^{\alpha(\mathbf{r})}$, where $b$ is a constant.

One could naively think that a direct one-to-one correspondence could exist between the spatial distribution of the NbN nano-crystals encoded in the STM topography and the gap map or
AA exponent map. In fact we found no cross-correlation between the local gap or exponent values and the topography of the probed area (see panel~\ref{Fig2}d). This is in agreement with our previous work \cite{Carbillet}. It could be partly due to the inhomogeneous oxidation of the NbN surface if this process would change the one-to-one correspondence between the NbN crystallites distribution and the STM topography. This nevertheless seems unlikely regarding the size of the nanostructures seen in STM which agree well with the size of the nanocrystals measured from TEM experiments.

In contrast, the present results furnish a new way of characterizing locally and quantitatively the underlying disorder distribution combined to electron interaction effects which lead to the appearance of supergrains of size $l_{\Delta}$. Indeed our results provide a new type of information that was not reported before, to our knowledge. For instance in Refs.\ \onlinecite{Sacepe2011} or \onlinecite{Kamlapure}, the focus of the STS studies is set on the comparison between local gap values and local peak height values.

We finish this experimental section by giving the numerical values of relevant microscopic parameters that will be used hereafter in our theoretical analysis. The low-temperature coherence length can be estimated as $\xi(0) = \sqrt{\hbar D/\Delta} \approx 5.2$ nm. It corresponds to the diffusion coefficient $D \approx 0.5$ cm$^2$/s obtained from $dB_{c2}(T)/dT$ data close to $T_c$,
see Ref.~\onlinecite{Semenov}. It is also consistent with our own electrical resistivity measurements in perpendicular high-magnetic field \cite{Carbillet} on similar films. Taking for the elastic mean free path $l \approx 0.5$ nm \cite{Semenov}, we find for the elastic scattering time: $\tau = l^2/3 D \approx 1.7 \times 10^{-15}$ s.

\section{Theoretical modeling}

Our theoretical explanation of the observed results is based on the standard understanding of:
\begin{itemize}
  \item[(i)] the Altshuler-Aronov anomaly \cite{Altshuler,Altshuler1985}, properly generalized beyond perturbation theory for both weak and strong coupling \cite{LevitovShytov,KamenevAndreev};
  \item[(ii)] the disorder-induced suppression of $T_c$ in superconducting films \cite{Finkelstein,Finkelstein1994}.
\end{itemize}
Both effects
have been described for a spatially homogeneous case. Both have thus to be properly generalized to an inhomogeneous situation, since this corresponds to the case of our experimental results.

The main physical mechanism leading to the suppression of $T_c$ in dirty superconductors is through the enhancement of Coulomb repulsion, which in turn reduces the effective attraction in the Cooper channel. An important distinction should here be made between the case of 2D and 3D superconductors.
In the 3D case, the whole effect can be simply described in terms of the Coulomb pseudopotential \cite{AMR}. However, our NbN ultrathin film is a 2D superconductor. Here, in contrast to the 3D situation, one cannot reduce the effect of Coulomb interaction to the modification of just a single Cooper coupling constant. In 2D one should instead track the whole energy dependence of the effective Cooper coupling, which changes considerably within the relevant energy scale of $T_c$. This requires a more sophisticated analysis, which can be done via a renormalization group technique developed by Finkelstein \cite{Finkelstein,Finkelstein1994}.

In 2D, the effects (i) and (ii) listed above are essentially determined by the film resistivity. Since we want to describe an inhomogeneous situation for these two effects, we assume a model, where the local resistivity $\rho(\br)$ fluctuates with the correlation length $l_\alpha$. With particular details relegated to Appendices, we summarize below the main findings.

\subsection{Link between the tunneling LDOS suppression and the local resistance}

According to Refs.\ \onlinecite{Altshuler,Altshuler1985,Ingold,LevitovShytov,KamenevAndreev}
the tunneling DOS suppression in an interacting diffusive normal metal, $\nu(E)$, is determined by the spreading resistance between the diffusive scale and the field propagation scale (see Appendix \ref{A:A1}).
Relevant $eV$ values (dozens of mV) are much smaller than $\hbar/\tau \approx 0.33$ eV, justifying well the diffusive regime.
In the 2D geometry, the log-normal dependence of
$\nu(E)$ \cite{LevitovShytov,KamenevAndreev} can be approximated by a power law with a weakly energy-dependent exponent
\be
\label{alpha}
  \alpha(V)
  =
  \frac{R_\Box}{2\pi R_Q}
  \ln \frac{\hbar\omega_0}{eV} ,
\ee
where $R_Q=h/e^2=25.8$ k$\Omega$ is the resistance quantum and
$\omega_0$ is the plasma frequency.
Taking $\hbar\omega_0\approx10$ eV  from
Ref.~\onlinecite{Semenov},  $V=10$ mV (typical voltage scale for experimental determination of $\alpha$, see Fig.\ \ref{Fig1}c) and $R_\Box=6.8$ k$\Omega$ (maximal sheet resistance $R_\Box^\text{max}$ before the transition, see Fig.\ \ref{Fig1}a), we obtain $\alpha=0.29$. This value is quite close to the mean value 0.35 measured experimentally. A small discrepancy could be related to the use of an underestimated $R_\Box^\text{max}$ that is significantly affected by thermal amplitude superconducting fluctuations, whereas the Altshuler-Aronov effect involves larger frequencies, where this suppression is less efficient. We conclude that the mean value of the AA exponent $\langle\alpha(\mathbf{r})\rangle$ is well described by the theory developed in Refs.\ \onlinecite{LevitovShytov,KamenevAndreev}.

The key point in our explanation of inhomogeneity seen in $\alpha(\mathbf{r})$ and $\Delta(\mathbf{r})$ is an assumption that they both originate from fluctuations of the local 2D resistivity $\rho(\mathbf{r})=R_\Box+\delta\rho(\mathbf{r})$.
Using the concept of the spreading resistance to calculate $\delta\alpha(\mathbf{r},V)$, we obtain
(see Appendix \ref{A:A2}):
\begin{equation}
\label{a-r}
  \delta\alpha(\mathbf{r},V)
  =
  \frac{\delta\rho(\mathbf{r})}{2\pi R_Q}
  \ln \frac{eV}{\hbar D/l_\alpha^2} .
\end{equation}
Using the experimentally measured dispersion of the AA exponent $\sigma_\alpha=0.028$ extracted from panel~\ref{Fig2}f, its correlation length $l_\alpha\approx18$~nm and taking $V=10$ mV, we find $eV/(\hbar D/l_\alpha^2) \approx 70 $ and hence obtain the sheet resistance dispersion $\sigma_\rho = 1.1$ k$\Omega$. This value corresponds to 16 \% of $R_\Box^\text{max}$. This dispersion value among local resistivities seems quite reasonable regarding the moderate but not too strong disorder existing in our NbN thin films.

We would like here to comment on the justification and meaning associated with the definition of a local resistivity. The correlation length $l_{\alpha}\approx18$~nm of the fluctuations of $\alpha$ exceeds well the electronic mean free path $l\approx0.5$~nm. This implies that our sample is deep in the diffusive regime for the considered energy scale corresponding to $\alpha$ and its variations,
thus the local resistivity can be defined. Since the spatial fluctuations of the AA exponent are mainly controlled by the diffusive length scale (the diffusive length $L$ is related to the energy $E$ through the relation $L=\sqrt{\hbar D/E}$), these fluctuations enable us to probe the local film resistivity.

\subsection{Link between the local gap inhomogeneities and the local resistances}

The additional suppression of the superconducting order parameter in regions with larger local resistivity is explained by the Finkelstein mechanism. In general, one should distinguish between spatial fluctuations coming from small ($q<q_D$) and large ($q>q_D$) momenta, where $q_D = \sqrt{\omega_D/D}$ with $\omega_D$ the Debye frequency. For our NbN film we estimate $\hbar\omega_D\approx 300$ K, $q_D\approx 1$ nm$^{-1}$. The low-momentum contribution is given by the usual log-cube 2D expression
\begin{equation}
\label{d-r}
  \frac{\delta\Delta(\mathbf{r})}{\left<\Delta\right>}
  \approx
  -
  \frac{\delta\rho(\mathbf{r})}{6\pi R_Q}
  \ln^3 \frac{\hbar\omega_D}{\Delta(0)} ,
\end{equation}
cut off at the Debye frequency (see Appendix \ref{SS:gap-homo}).
The high-momentum contribution can be expressed in terms of the renormalization of the BCS coupling constant $\lambda(\textbf{r})$.
Substituting the resistivity dispersion $\sigma_\rho = 1.1$ k$\Omega$ into Eq.\ \eqref{d-r}, we obtain for the relative gap dispersion $\sigma_\Delta/\left<\Delta\right> = 0.074$, whereas the experimental value is 0.057 (see panel~\ref{Fig2}e). Note that Eq.\ \eqref{d-r} contains a large degree of uncertainty: the precise value of $\hbar\omega_D$ being unknown, $\sigma_\Delta$ can only be determined with roughly 50 \% accuracy.

Nevertheless, the fact that Eq.\ \eqref{d-r} nearly describes the measured gap dispersion indicates that the high-momentum contribution to spatial fluctuations of $\Delta(\mathbf{r})$ is absent. This case could physically correspond to a situation, where the 2D resistivity would spatially fluctuate due to the variations of interface resistance between neighboring crystallites (that would lead to small $q<q_D$ momenta contribution), while the 3D resistivity within each crystallite would remain identical (no change in the contribution of the $q>q_D$ momenta).

\section{Discussion}
We have shown that local probe spectroscopic studies are able to demonstrate a direct relation between the properties of the electronic excitation spectrum pertinent to the lowest-energy scale: the superconducting gap ($\sim1$ meV) on one hand, and to much higher energy scales (few dozens of meV) relevant to the Coulomb anomaly induced by the Altshuler-Aronov effect on the other hand. This supports the relevance of the ``fermionic" mechanism and Finkelstein theory for superconductivity suppression by moderate disorder in NbN thin films not only at the global scale but also locally.

We characterized the nanoscale low-temperature granularity that develops in the superconducting NbN film and have shown that electron-electron interactions play a crucial role in determining this precise granularity. A natural generalization then would be to study and understand the temperature evolution of the inhomogeneity of $\Delta(\textbf{r})$ and $\alpha(\textbf{r})$, in particular around and above $T_c$, to connect to the pseudogap phenomenon and its current understanding in various materials. Indeed, we point out that for the present ultrathin film the ratio $\Delta(0)/k_BT_c \approx 3.7$ is twice larger than the BCS theory prediction 1.76, and significantly larger than the value 2.1 known for thicker and less disordered NbN films \cite{Semenov}, fabricated using the same set-up, procedures and recipees than in the present work.

The increase of the $\Delta(0)/k_BT_c$ ratio with the sheet resistance can be understood within the usual mechanism of strong thermal phase fluctuations~\cite{EmeryKivelson}, which are known to be important in 2D superconducting systems with small superfluid density (suppressed due to strong
disorder). Indeed, the width of the Gaussian thermal fluctuational region in 2D, as seen from conductivity measurements, can be estimated by comparing the Aslamazov-Larkin paraconductivity $\sigma_\text{AL}^\text{2D} = (e^2/16\hbar)\ln(T/T_{c0})$ and normal-state conductivity as $\mathrm{Gi} = (e^2/16\hbar)R_\Box $, where $T_{c0}$ is the mean-field transition temperature.
Taking into account that phase fluctuations are about three times stronger than Gi\cite{LV-book}, we obtain an estimate for the renormalized $T_c$, which coincides with the Berezinskii-Kosterlitz-Thouless transition: $(T_{c0}-T_{c})/T_{c0} \approx 3\, \text{Gi}$.
With the normal-state resistance $R_\Box \approx 7$ KOhm (see Fig.\ \ref{Fig1}a), this gives $T_{c0} \approx 1.4 T_{c}$.
Thus we conclude that the major part of the increase of the $\Delta(0)/k_BT_c$ ratio can be understood as being due to fluctuation suppression of the measured $T_c$ with respect to the mean-field value $T_{c0}$, which should be compared to the gap width $\Delta(0)$.  The same conclusion was reached in Ref.\ \onlinecite{Chand2012}, where much thicker (actually 3D) NbN films were studied.

The above explanation of the enhanced $\Delta(0)/k_BT_c$ and pseudogap phenomenon (discussed in more details in the previous paper~\cite{Carbillet}) refers to the ``bosonic" mechanism of phase fluctuations, while local correlation between Altshuler-Aronov exponent $\alpha(\textbf{r})$ and the low-temperature gap value $\Delta(\textbf{r})$ are nicely explained within the ``fermionic" mechanism.  In that sense NbN films seem to demonstrate simultaneous action of \textit{different}
superconductivity suppression mechanisms. It would be interesting to see if strongly disordered NbN films exhibit a double-gap behavior found recently in InO$_x$ films\cite{Dubouchet2019}.
Other methods such as the ones addressing local coherence through scanning Josephson microscopy \cite{Bergeal2008} by mapping the local critical current might also shed important new light \cite{Cho2019}, when compared to other local
spectroscopic properties such as the local gap map or peak height map. Finally, addressing the LDOS measurements and its mesoscopic fluctuations near $T_c$ in disordered thin films similar to ours would also help testing and comparing with recent theoretical predictions\cite{SF2005,we2012,Burmistrov}.

\section{Conclusion}

In conclusion, we have demonstrated sharp spatial anti-correlations in highly resistive NbN thin films between the local magnitude of the superconducting energy gap $\Delta(\textbf{r})$ and the value of the exponent $\alpha(\textbf{r})$ characterizing the local density-of-states suppression at the Fermi level due to the Altshuler-Aronov effect induced by Coulomb repulsion. The correlation lengths characteristic of the emergent inhomogeneities at $T \ll T_c$ are $l_{\Delta} \approx 27$~nm for $\Delta(\textbf{r})$ and $l_{\alpha} \approx 18$~nm for $\alpha(\textbf{r})$, while the cross-correlation length $l_\text{cross}$ between $\Delta(\textbf{r})$ and $\alpha(\textbf{r})$ is $l_\text{cross} \approx 25$~nm. Our results are in agreement with the predictions of a local version of the Finkelstein theory for the ``fermionic" mechanism of superconductivity suppression by disorder. This behavior is generally expected for any 2D superconducting materials satisfaying: (a) a strong Coulomb interaction between conduction electrons, (b) a spatial scale $l_{\alpha}$ characteristic of the inhomogeneities of the Altshuler-Aronov density-of-states suppression much larger than the superconducting coherence length $\xi(0)$. Our results suggest that in NbN films the emerging granularity can be fully described by a ``fermionic" mechanism. Additional theoretical and experimental work is needed to see if the same claim is valid for other recently discovered 2D disordered materials, including monolayer or bilayer dichalcogenides \cite{Costanzo, Xi}, bilayer graphene \cite{Cao} or interface oxydes \cite{Caprara}.

\acknowledgments

We thank D. S. Antonenko for useful discussions. This work was supported by the French ANR project SUPERSTRIPES under the reference number ANR-15-CE30-0026, by the Russian Scientific Foundation under Grant No.\ 20-12-00361 and by the Russian Academy of Sciences Program ``Low temperature physics".

\appendix

\section{Numerical computations of the auto-correlation and cross-correlation functions}
\label{correl}

The auto-correlation functions calculated from the gap map presented in Fig.\ \ref{Fig2}b and the exponent map in Fig.\ \ref{Fig2}c are computed using the 2D expression (\ref{autocorr}) with periodic boundary conditions to treat the finite size spectroscopic data. We chose such periodic boundary conditions for simplicity. Nevertheless, these latter conditions do not affect the global behavior of the auto-correlation function since the correlation lengths that are extracted from the dependence of the auto-correlation functions are about $1/10-1/20$ of the image size. Thus the spatial dependence of the auto-correlation function is not affected by boundary conditions in our case.

In order to ensure proper normalization, the auto-correlation functions are calculated by subtracting the mean value of the variable of interest and by dividing by the standard deviation $\sigma$ of this variable. For instance for the local gap values $\Delta(\mathbf{r})$ the 2D auto-correlation function $\rho_{\Delta}(\mathbf{r})$ is defined by
\be
\label{autocorr}
	\rho_{\Delta}(\br) =  \sum_{\br'} \frac{(\Delta(\br') - \corr{\Delta}) (\Delta(\br' - \br) - \corr{\Delta})}{N \sigma_{\Delta}^2} ,
\ee
where $N$ is the total number of pixels in the $\Delta(\mathbf{r})$ gap map. The 2D auto-correlation functions were calculated in this way for both $\Delta(\mathbf{r})$ and $\alpha(\mathbf{r})$ data presented in Fig.\ \ref{Fig2}. Since $\rho_{\Delta}(\br)$ and $\rho_{\alpha}(\br)$ were rather isotropic we decided to circularly average them to extract the correlation lengths $l_{\Delta}$ and $l_{\alpha}$. The result of the circular averaging of $\rho_{\Delta}(\br)$ and $\rho_{\alpha}(\br)$ is presented in Fig.\ \ref{Fig3}, where the spatial dependence of $\rho_{\Delta}(r)$ and $\rho_{\alpha}(r)$ is plotted as a function of the distance $r$. The correlation lengths are determined in such a way that the correlation functions remain above the noise level close to zero.

The 2D cross-correlation function $\rho_\text{cross}(\br)$ between $\Delta(\mathbf{r})$ and $\alpha(\mathbf{r})$ is calculated in the same way as above using also periodic boundary conditions with the following formula:
\be
\label{crosscorr}
	\rho_\text{cross}(\br) =  \sum_{\br'} \frac{(\Delta(\br') - \corr{\Delta}) (\alpha(\br' - \br) - \corr{\alpha})}{N \sigma_{\Delta} \sigma_{\alpha}} .
\ee
$\rho_\text{cross}(\br)$ is plotted in panel g of Fig.\ \ref{Fig2}. The 2D cross-correlation function between $\Delta(\br)$ and the topography $Z(\br)$ presented in panel a of Fig.\ \ref{Fig2} is calculated in the same way, $\alpha(\br)$ in (\ref{crosscorr}) being replaced by $Z(\br)$. Finally since $\rho_\text{cross}(\br)$ has a rather isotropic behavior, as seen on Fig.\ \ref{Fig2}g, it was circularly averaged in order to extract the cross-correlation length $l_\text{cross}$ from the spatial dependence of $\rho_\text{cross}(r)$.

\section{Altshuler-Aronov anomaly in inhomogeneous films}
\label{A:A}

We consider the tunneling density of states $\nu(E)$ in a diffusive normal film, where the local resistivity fluctuates with the correlation length $a$. The mean free path $l$, the film thickness $d$ and other parameters satisfy the experimentally relevant set of inequalities:
$l \ll d \ll \sqrt{D/E} \ll a$.
Here and below we use the system of units with $k_B=\hbar=1$.

\subsection{Homogeneous film}
\label{A:A1}

The theory of the Altshuler-Aronov anomaly, also called zero-bias anomaly, \cite{Altshuler,Altshuler1985,Ingold,Fin1983,LevitovShytov,KamenevAndreev}
predicts that the Coulomb-induced suppression of the tunneling density of states can be written in the form $\nu(E)=\nu_0 e^{-S(E)}$, where $\nu_0$ is the bare density of states at the Fermi level in the absence of interaction. The function $S(E)$ can be considered as the gauge phase correlation function \cite{Fin1983,KamenevAndreev} or the instanton action \cite{LevitovShytov}. In the case of diffusive electron motion it can be written in the form
\be
\label{S-int}
  S(E) =
  \frac{2}{R_Q}
  \int_E^{1/\tau}
  \frac{d\omega}{\omega} \, R(\omega) ,
\ee
where $R_Q = h/e^2$ is the resistance quantum and
$R(\omega)$ is the spreading resistance between the diffusive length $r_\text{in}(\omega)=\sqrt{D/\omega}$ and electromagnetic field propagation length $r_\text{out}(\omega)=\sqrt{D\omega_0/\omega^2}$, where $\omega_0$ is the plasma frequency.
In the 2D geometry with a constant sheet resistance $R_\Box$,
\be
  R(\omega)
  =
  \frac{R_\Box}{2\pi}
  \ln \frac{r_\text{out}(\omega)}{r_\text{in}(\omega)}
  =
  \frac{R_\Box}{4\pi}
  \ln \frac{\omega_0}{\omega} .
\ee
Taking the integral in Eq.\ (\ref{S-int}) we arrive at the standard result for homogeneous films:
\be
\label{S-res-homo}
  S_\text{homo}(E)
  =
  \frac{R_\Box}{4\pi R_Q}
  \left(
  \ln^2 \frac{\omega_0}{E}
  -
  \ln^2 \omega_0\tau
  \right) .
\ee
For practical purposes, the slow log-normal energy dependence of $\nu(E)$ predicted by Eq.\ \eqref{S-res-homo} can be replaced by the power law $\nu(E)\propto E^\alpha$, with an almost energy-independent index $\alpha(E) = - \partial S_\text{homo}(E)/\partial\ln E$, leading to Eq.\ \eqref{alpha}.

\subsection{Inhomogeneous film}
\label{A:A2}

Consider now the Altshuler-Aronov anomaly in an inhomogeneous film. We will treat inhomogeneity in a simple model, where the local sheet resistance can be written as $\rho(\br) = R_\Box + \delta\rho(\br)$, with $\delta\rho(\br)$ being a Gaussian noise with zero average and the correlation length $a$.
The spreading resistance $R(\br_0,\omega)$ now depends on the tunneling position $\br_0$ and can be estimated using logarithmic dependence of resistance in 2D. Therefore, with logarithmic accuracy one may calculate $R(\br_0,\omega)$ replacing $\rho(\br)$ by $\rho(\br_0)$ for $|\br-\br_0|<a$ and by $R_\Box$ for $|\br-\br_0|>a$.

In the experimentally relevant case, $E\gg D/a^2$, and therefore $r_\text{in}(\omega)\ll a$. Then the spreading resistance $R(\br,\omega)$ can be estimated as
\be
  R(\br,\omega)
  =
  \frac{\rho(\br)}{2\pi}
  \ln \frac{a}{r_\text{in}(\omega)}
  +
  \frac{R_\Box}{2\pi}
  \ln \frac{r_\text{out}(\omega)}{a}
\ee
Substituting that into Eq.\ \eqref{S-int}, we get
\be
\label{S-int-loc}
  S(\br,E)
  =
  S_\text{homo}(E)
  +
  \frac{\delta \rho(\br)}{4\pi R_Q}
  \left(
    \ln^2 \frac{1/\tau}{D/a^2}
    -
    \ln^2 \frac{E}{D/a^2}
  \right) ,
\ee
where $S_\text{homo}(E)$ is given by Eq.\ \eqref{S-res-homo}.
Taking the derivative, we see that the AA exponent fluctuates according to Eq.\ \eqref{a-r},
where we identify $a$ with $l_\alpha$.

\section{Gap fluctuations in inhomogeneous films}

If the correlation length of the local resistivity fluctuations exceeds the superconducting coherence length, $a\gg\xi(0)$, then the spectral gap at a given point $\br$ is determined by the order parameter $\Delta$ in a homogeneous system with the resistivity $\rho=\rho(\br)$. The dependence of $\Delta$ on $\rho$ results from the Coulomb repulsion enhanced in a diffusive media \cite{Altshuler1985}. Assuming the effect can be treated perturbatively, we will rederive the known expression \cite{Ovchina,Maekawa,Takagi,Fukuyama} focusing on proper cutoffs of energy and momentum integrals.

\subsection{Order parameter suppression in homogeneous films}
\label{SS:gap-homo}

Coulomb suppression of superconductivity \cite{Fin1983} can be described in terms of the energy-dependent correction to the Cooper channel interaction constant \cite{we2012}:
\be
\label{d-lambda}
  \delta\lambda_{\eps,\eps'}
  =
  - \frac{1}{2\pi\nu}
  \int (dq)
  \frac{1}{Dq^2+\eps+\eps'} ,
\ee
where $\nu$ is the 3D density of states at the Fermi energy per one spin projection, both Matsubara energies, $\eps$ and $\eps'$, are assumed to be larger than $T_c$, and
\be
\label{int-sum}
  \int (dq) f(q^2)
  \equiv
  \frac1d \int \frac{dq_x dq_y}{(2\pi)^2} \sum_{n=0}^\infty
  f[q_x^2+q_y^2+(\pi n/d)^2] .
\ee
Perturbative account of $\delta\lambda_{\eps,\eps'}$ in the normal state reproduces the log-cube shift of $T_c$ \cite{Ovchina,Maekawa,Takagi,Fukuyama}. In the superconducting state, it leads to the shift of the order parameter compared to the clean case. In the low-temperature limit ($T\ll T_c$), this shift is given by \cite{we2012}
\be
\label{dd/d}
  \frac{\delta\Delta}{\Delta_0}
  =
  \int_\Delta^{\omega_D}
  \frac{d\eps}{\eps}
  \frac{d\eps'}{\eps'}
  \,
  \delta\lambda_{\eps,\eps'} ,
\ee
where $\Delta_0$ is the gap value in the clean system.
This formula generalizes the BCS relation $\delta\Delta/\Delta=\delta\lambda/\lambda^2$ to the case of an energy-dependent coupling constant.

Since the energies $\eps$ and $\eps'$ in Eq.\ \eqref{dd/d} are limited by $\omega_D$, one can represent the general expression (\ref{d-lambda}) as
\be
\label{dlambda2}
  \delta\lambda_{\eps,\eps'}
  \approx
  \delta\lambda_{\eps,\eps'}^\text{low}
  +
  \delta\lambda^\text{high} ,
\ee
where $\delta\lambda_{\eps,\eps'}^\text{low}$ is the energy-dependent contribution from low ($q<q_D$) momenta and $\delta\lambda^\text{high}$ is the energy-independent contribution from high ($q>q_D$) momenta:
\begin{gather}
\label{dlambdas}
  \delta\lambda_{\eps,\eps'}^\text{low}
  =
  - \frac{1}{2\pi\nu}
  \int (dq)
  \frac{\theta(q_D-q)}{Dq^2+\eps+\eps'},
\\{}
  \delta\lambda^\text{high}
  =
  - \frac{1}{2\pi\nu}
  \int (dq)
  \frac{\theta(q-q_D)}{Dq^2},
\end{gather}
and $q_D = \sqrt{\omega_D/D}$ is determined by the Debye frequency.

The high-momentum contribution $\delta\lambda^\text{high}$ is just a correction to the Cooper channel constant. According to Eq.\ \eqref{dd/d} it leads to the perturbative suppression of the order parameter: $\delta\Delta^\text{high}/\Delta=\delta\lambda^\text{high}/\lambda^2$.
This expression can be extended beyond the perturbation theory using the BCS relation that results in the renormalization of the bare gap: $\Delta_0 \to \Delta_1$, where \be
\label{Delta1}
\Delta_1 = \Delta_0 \exp[1/\lambda-1/(\lambda+\lambda^\text{high})].
\ee
Once the high-momentum contribution is incorporated into $\Delta_1$, the perturbative shift of the gap due to the low-momentum part $\delta\lambda^\text{low}_{\eps,\eps'}$ should be determined by Eq.\ \eqref{dd/d} with $\Delta_0$ replaced by $\Delta_1$.

In order to estimate the low-momentum contribution $\delta\lambda^\text{low}_{\eps,\eps'}$ for our NbN films, we have to compare $\omega_D\approx 300$ K and the transverse Thouless energy $\hbar D(\pi/d)^2\approx 950$ K. Since the latter is larger, one can leave only the $n=0$ mode in Eq.\ \eqref{int-sum}, leading to the 2D contribution (the 2D resistivity is $\rho = R_Q/4\pi \nu D d$)
\be
  \delta\lambda_{\eps,\eps'}^\text{low}
  =
  - \frac{\rho}{2\pi R_Q} \ln \frac{\omega_D}{\max(\eps,\eps')}
\ee
and hence to the following shift of $\Delta$:
\be
\label{ddl}
  \frac{\delta\Delta^\text{low}}{\Delta_1}
  =
  - \frac{\rho}{6\pi R_Q}
  \ln^3 \frac{\omega_D}{\Delta} .
\ee

For our NbN films with $\rho=R_\Box^\text{max} = 6.8$ k$\Omega$, $\hbar\omega_D\approx300$ K and $\Delta=12$ K, the low-momentum contribution to the gap suppression according to Eq.\ \eqref{ddl} is given by $\delta\Delta^\text{low}/\Delta_1 \approx 0.55$, which is at the border of applicability of the perturbation theory.
At the same time, $\lambda^\text{high}$ in Eq.\ (\ref{dlambdas}) is determined by
the 3D diffusion which should be cut off at $q\sim1/l$:
$\delta\lambda^\text{high} \sim - (\rho/\pi^2R_Q)(d/l) \sim 0.1$, yet with an unknown prefactor. Depending on the latter, $\Delta_1/\Delta_0$ is in the range of several tens of percent, that is a bit too small to describe experimental data.

\color{black}
\subsection{Inhomogeneous films}

In Sec.\ \ref{SS:gap-homo}, we calculated the gap suppression $\delta\Delta$ with respect to the clean case ($\rho\to0$). To get gap fluctuations $\delta\Delta(\br)$ in an inhomogeneous system with long-range fluctuations of $\rho(\br)$, we simply replace $\rho$ by $\delta\rho(\br)$:
\be
\label{de-fluct}
  \frac{\delta\Delta(\br)}{\corr{\Delta}}
  =
  - \frac{\delta\rho(\br)}{6\pi R_Q}
  \ln^3 \frac{\omega_D}{\Delta}
  +
  \delta\lambda^\text{high}(\br)
  \ln^2 \frac{\omega_D}{\Delta}
  ,
\ee
where $\delta\lambda^\text{high}(\br)$ contains the large-momenta contribution:
\be
  \delta\lambda^\text{high}(\br)
  =
  -
  \frac{2d\,\delta\rho(\br)}{R_Q}
  \int (dq)
  \frac{\theta(q-q_D)}{q^2} .
\ee

Equation \eqref{de-fluct} is derived in the model, where the resistivity [both the 3D $\rho_3(x,y)$ and the 2D $\rho(x,y)$ related via $\rho_3=\rho/d$] exhibits small spatial fluctuations. In a more realistic model presumably applicable to our NbN films, these are the crystalline interfaces that are the sources of resistance fluctuations at scales larger that a few nm. For such a model of structural disorder, high-momentum fluctuations of $\delta^\text{high}(\br)$ are absent, and one arrives at Eq.\ \eqref{d-r}.




\end{document}